\documentclass[
    reprint, 
    amsmath, 
    amssymb, 
    aps, 
    prl, 
    floatfix, 
    superscriptaddress
]{revtex4-2}
\pdfoutput=1

\usepackage{soul}
\usepackage[T1]{fontenc} 
\usepackage{lmodern}
\usepackage{amsmath}
\usepackage{array}
\usepackage{dsfont}
\usepackage{mathtools}
\usepackage{graphicx}
\usepackage{dcolumn}
\usepackage{bm}
\usepackage{hyperref}
\usepackage{physics}
\usepackage{empheq}
\usepackage{color}
\usepackage[dvipsnames]{xcolor}
\usepackage{multirow}
\usepackage{orcidlink}
\usepackage{afterpage}
\usepackage{mathrsfs}
\usepackage{tikz}
\usepackage{ulem}
\usetikzlibrary{shapes}
\usetikzlibrary{tikzmark,calc}

\newcommand{\bs}[1]{\boldsymbol{#1}}

\allowdisplaybreaks

\begin{document}

\title{
    Superconductivity in the Metallic Wigner Crystals of Rhombohedral Graphene:\\ 
    Possible Realization of an Electronic Supersolid
}
%%%%%%%%%
\author{F\'elix Desrochers \orcidlink{0000-0003-1211-901X}}
\email{felix\_desrochers@fas.harvard.edu}
\affiliation{%
Department of Physics, Harvard University, Cambridge, MA 02138, United States
}%
\author{Mark R. Hirsbrunner\orcidlink{0000-0001-8115-6098}}
\affiliation{%
Department of Physics, University of Toronto, Toronto, Ontario M5S 1A7, Canada
}%
\author{Yong Baek Kim\orcidlink{0000-0003-0473-8785}}
\affiliation{%
Department of Physics, University of Toronto, Toronto, Ontario M5S 1A7, Canada
}%
%%%%
\author{Ashvin Vishwanath \orcidlink{0000-0002-6306-2263}}
\email{avishwanath@g.harvard.edu}
\affiliation{%
Department of Physics, Harvard University, Cambridge, MA 02138, United States
}%
\date{\today}

\begin{abstract}
Recent experiments on rhombohedral multilayer graphene have reported metallic Wigner crystals (MWCs), in which spontaneously broken translational symmetry coexists with mobile charge carriers. Their proximity to time-reversal-symmetry-breaking superconductivity raises the intriguing possibility that collective excitations of the electron crystal mediate pairing among the residual itinerant carriers, giving rise to a superconducting Wigner crystal---an electronic analog of a supersolid. Here we investigate this possibility using a realistic model of electron-doped rhombohedral multilayer graphene with spin and valley polarization, whose low symmetry removes the usual Cooper instability. We use self-consistent and time-dependent Hartree–Fock theory to map the evolution of the self-doping fraction and characterize the collective-mode spectrum of the MWCs, followed by a multiband linearized Eliashberg analysis of pairing. Upon increasing density, we find an instability of the MWC, triggered by a weakly dispersing unoccupied band approaching the chemical potential. Pairing remains weak deep in the stable MWC. However, on approaching the instability, we identify a collective mode composed of interband electron-hole excitations that softens. We show that this mode drives pairing at experimentally relevant temperatures. The leading pairing instability is at finite momentum and combines Fermi surface states with states in the incipient band, making a Fermi surface or single-band pairing analysis inadequate. Our results establish collective-mode-mediated superconductivity in MWCs as a viable scenario for rhombohedral graphene and motivate experimental searches for coexisting electronic charge order in its superconducting phases.
%Recent experiments on rhombohedral multilayer graphene have revealed evidence for metallic Wigner crystals (MWCs), in which spontaneously broken translational symmetry coexists with a non-integer number of electrons per emergent unit cell. These unusual states host mobile charge carriers alongside a dynamical periodic charge modulation. Their proximity to time-reversal-symmetry-breaking superconductivity raises the intriguing possibility that collective excitations of the electron crystal mediate pairing among the residual itinerant carriers, giving rise to a superconducting Wigner crystal---an electronic analog of a supersolid. Here we investigate this possibility using a realistic model of electron-doped rhombohedral multilayer graphene. We map the evolution of the self-doping fraction and characterize the collective-mode spectrum of the MWCs, identifying a soft collective mode that ultimately leads to an instability of the electron crystal. Importantly, we find that this soft collective mode drives a finite-momentum interband pairing instability at experimentally relevant temperatures, but only in close proximity to the electron crystal's instability. Our results establish collective-mode-mediated superconductivity in MWCs as a viable scenario for rhombohedral graphene and motivate experimental searches for coexisting electronic charge order in its superconducting phases.
\end{abstract}

\maketitle

\begin{figure*}[t]
    \centering
    \includegraphics[width=1.0\textwidth]{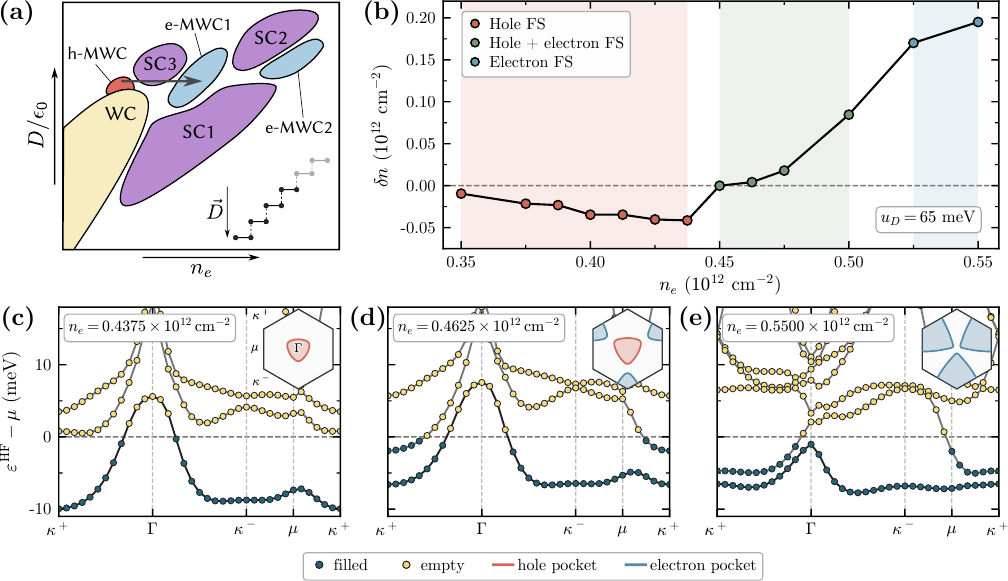}\vspace{-2mm}
    \caption{
    \textbf{Self-doping and Fermi-surface evolution in electron crystals.} 
    (a) Schematic density--displacement-field phase diagram showing WC, h-MWC, e-MWC, and superconducting regimes for R(4,5,6)G. The grey arrow denotes pictorially the regime we focus on. (b) Net self-doping $\delta n$, measured relative to one electron per crystal cell, for the lowest-energy converged HF states among the sampled dopings at $u_D=65\,\mathrm{meV}$ for R4G. Colors distinguish hole-only (red), coexisting electron--hole (green), and electron-only (blue) Fermi surfaces. (c--e) Representative HF dispersions and Fermi-surface insets for these three regimes, respectively. Marker colors distinguish occupied and empty states. Energies are measured relative to the chemical potential.
    }
    \label{fig:1_HF_self-doping}
\end{figure*}

%%%%%%%%%%%%%%%%
% Introduction %
%%%%%%%%%%%%%%%%
\textit{Introduction}.
Rhombohedral multilayer graphene (RMG) has emerged as a versatile platform for exploring strongly correlated electronic phases, owing to the enhanced Coulomb interactions and concentrated Berry curvature in its highly tunable band structure. Recent experiments have revealed a remarkable variety of correlated states, including integer and fractional quantum anomalous Hall phases~\cite{lu2024fractional, lu2025extended, xie2025tunable, waters2025chern, aronson2025displacement, Su2025, huo2025does, uzan2025hbn, zhang2025moire, li2025stacking, liu2026odd}, electron crystals~\cite{hanEvidenceMetallicWigner2026, nguyenHierarchySuperconductivityTopological2025, hanSignaturesChiralSuperconductivity2025a, zhouCompetingOrdersDriven2026a, nguyenCoexistingChargeDensity2026,sunRealspaceVisualizationEmergent2026}, and unconventional superconductivity (SC) that spontaneously breaks time-reversal symmetry~\cite{hanSignaturesChiralSuperconductivity2025a, choi2025superconductivity, yang2025magnetic, seo2025family, dutta2026reconfigurable, kalantreFermiologyCandidateChiral2026a, zhengChiralSuperconductorsCompeting2026}. Particularly intriguing is the observation of metallic electron crystals, in which translational symmetry is spontaneously broken while a finite density of itinerant charge carriers remains~\cite{hanSignaturesChiralSuperconductivity2025a, nguyenHierarchySuperconductivityTopological2025, hanEvidenceMetallicWigner2026, nguyenCoexistingChargeDensity2026, zhouCompetingOrdersDriven2026a}. Unlike conventional insulating Wigner crystals (WCs), these metallic Wigner crystals (MWCs) typically possess a non-integer number of electrons per emergent unit cell, allowing crystalline charge order and metallic transport to coexist~\cite{dongCrystalsCaughtDoping2026, fengSelfdopedCrystalPreempted2026, kim2026shape, geNematicWignerCrystals2026a, abouelkomsan2026rhombohedral, guerciQuantumOscillationsHole2026}.

A distinctive feature of MWCs is the coexistence of itinerant carriers with low-energy collective excitations associated with the crystalline electronic order. In particular, the phonon modes of the electron crystal provide a natural mechanism for generating effective attractive interactions between the residual mobile carriers. Were these interactions sufficiently strong to induce superconductivity, the resulting state would simultaneously exhibit spontaneous translational symmetry breaking and superconducting phase coherence. Such a phase---a superconducting WC---would constitute an electronic analog of a supersolid~\cite{leggett1970can, chester1970speculations, andreev1971quantum, anderson2005thermodynamics, prokof2005supersolid, RevModPhys.84.759}. The experimentally observed proximity between metallic electron crystals and time-reversal-symmetry-breaking superconductivity in RMG makes this possibility especially compelling. Importantly, however, broken time-reversal symmetry can remove the symmetry protection of the conventional weak-coupling Cooper instability, and trigonal warping generates an intravalley asymmetry between opposite momenta that can further suppress pairing~\cite{fischer2018superconductivity, fischer2023superconductivity}. Establishing whether superconductivity can emerge within a self-consistent description of a metallic electron crystal in a realistic microscopic model of RMG thus remains an important open question, despite growing theoretical interest in related pairing mechanisms~\cite{miaoCooperInstabilityMagnetic2026, wangChiralSuperconductivityAdjacent2026}. 

In this work, we investigate this possibility using a realistic microscopic model of electron-doped rhombohedral multilayer graphene~\cite{auerbach2025isospin, dongCrystalsCaughtDoping2026}. We first employ self-consistent Hartree–Fock calculations to characterize the metallic electron crystal states over an experimentally relevant parameter regime. We identify three distinct regimes as we vary the carrier density: a hole-doped metallic Wigner crystal (h-MWC) with a hole pocket around the $\Gamma$ point at low carrier density, an intermediate regime in which electron and hole pockets coexist, and an electron-doped metallic Wigner crystal (e-MWC) with an electron pocket around the $\kappa_+$ point at high carrier density. We then use time-dependent Hartree–Fock (TDHF) theory to determine the stability of the crystalline states and obtain their collective excitations. We find a collective instability of the state once the electron pocket is nucleated. This instability is preceded by the progressive softening of a collective mode near the boundary of the Brillouin zone, which may drive a strong effective interaction between the itinerant carriers as the instability onsets.
%We find that the low-lying collective mode softens near the boundary of the Brillouin zone as the density increases towards the nucleation of the electron pocket, potentially mediating a strong effective interaction between the itinerant carriers.

To study pairing, we consider the retarded interaction generated by all TDHF charge-neutral excitations, in addition to the instantaneous Coulomb repulsion, and solve the multiband linearized Eliashberg equation using the HF propagator. We find that the mere coexistence of mobile carriers and electron-crystal collective modes does not necessarily yield appreciable superconducting pairing. Deep within the h-MWC phase, pairing around the Fermi surface remains weak, as expected from the suppression of the conventional Cooper instability due to the lack of inversion symmetry about the valley center. In contrast, we do find that the soft modes that appear prior to the crystal instability mediate interaction strong enough to lead to a large pairing instability. The nature of the soft mode further favors an unconventional superconducting instability involving finite-momentum interband pairing, despite the presence of a single Fermi surface in the lowest band. We establish the central role of this mode by showing that its contribution alone, together with the instantaneous Coulomb repulsion, is sufficient to sustain the pairing instability, whereas removing it suppresses superconductivity altogether. Our results establish a microscopic mechanism by which the collective fluctuations of an electron crystal can promote superconductivity, highlighting the potential for these seemingly competing orders to coexist in RMG.

%%%%%%%%%%%%%%%%%%%%%%%%%%%%%%%%
% Model and metallic Wigner crystal %
%%%%%%%%%%%%%%%%%%%%%%%%%%%%%%%%
\textit{Model and metallic Wigner crystal}.---In the low electron doping, large displacement field regime of interest, many experiments report similar phenomenology across R4G, R5G, and R6G, with phase diagrams broadly resembling the one depicted in Fig.~\ref{fig:1_HF_self-doping}(a). These phase diagrams exhibit insulating WC phases, various distinct superconducting pockets, and electron- and hole-doped metallic electron crystals~\cite{hanEvidenceMetallicWigner2026, nguyenHierarchySuperconductivityTopological2025, hanSignaturesChiralSuperconductivity2025a, zhouCompetingOrdersDriven2026a, nguyenCoexistingChargeDensity2026,sunRealspaceVisualizationEmergent2026, hanSignaturesChiralSuperconductivity2025a, choi2025superconductivity, yang2025magnetic, seo2025family, dutta2026reconfigurable, kalantreFermiologyCandidateChiral2026a, zhengChiralSuperconductorsCompeting2026}~\footnote{We note that evidence for individual phases, especially the electron-doped crystal, varies with layer number. In particular, evidence for e-MWC has been reported only in 5-layer~\cite{zhouCompetingOrdersDriven2026a} and 6-layer~\cite {nguyenCoexistingChargeDensity2026} devices.} For concreteness, we focus on rhombohedral tetralayer graphene (R4G) throughout this work. We utilize a continuum model for interacting R4G~\cite{PhysRevB.82.035409},
\begin{align}
    \hat{H}=\hat{h}_{\mathrm{R4G}} + \frac{1}{2 A} \sum_{\bs{q}} V_{\bs{q}}: \hat{\rho}_{\bs{q}} \hat{\rho}_{-\bs{q}}:, \label{eq:continuum_hamiltonian_rng}
\end{align}
where $A$ is the system area, $\hat\rho_{\bs{q}}$ is the density operator at momentum $\bs{q}$, and normal ordering is taken relative to the charge-neutral gap of R4G. The single-particle Hamiltonian, $\hat h_{\rm R4G}$, is a standard tight-binding model for rhombohedral graphene (see Supplemental Material~\footnote{See Supplemental Material.} for more details), and we model the perpendicular displacement field as a potential difference $u_D$ between adjacent layers. We consider a dual-gated screened Coulomb interaction $V_{\bs{q}} = e^2 \tanh (q d_s)/(2 \epsilon_r \epsilon_0 q)$ using a gate spacing $d_s = 25$~nm and a relative permittivity $\epsilon_r = 15$, the latter of which is chosen following Refs.~\cite{dongCrystalsCaughtDoping2026, auerbach2025isospin}.

We perform self-consistent Hartree-Fock (HF) calculations on Eq.~\eqref{eq:continuum_hamiltonian_rng} assuming polarization to a single spin and valley. To capture crystallization, we restrict the HF ans\"atze to states invariant under the discrete translations of a triangular lattice. The lattice constant $L \sim 15$~nm is treated as a variational parameter that sets the unit-cell density $n_{\rm u.c.} = 2/(\sqrt{3}L^2)$, which we allow to differ from the electron density $n_e$ to accommodate incommensurate crystals. We project onto the $N_b = 7$ conduction bands closest to the Fermi level and use $N_k = 36 \times 36$ unit cells throughout. The HF state is characterized by the one-body reduced density matrix $\mathcal{P}_{ab}(\bs{k}) = \langle \hat c^\dagger_{\bs{k}a} \hat{c}_{\bs{k}b} \rangle$, where $a,b$ label zone-folded states of the projected parent bands and $\bs{k}$ is the reduced momentum in the crystalline Brillouin zone (BZ). The HF Hamiltonian is $\hat{h}_{\rm HF}[\mathcal{P}] = \hat{h}_{\rm R4G} + \hat{h}_H[\mathcal{P}] + \hat{h}_F[\mathcal{P}]$, with Hartree and Fock terms $\hat{h}_H$ and $\hat{h}_F$, is iterated to self-consistency~\cite{Note2}, yielding reconstructed bands $\varepsilon^{\rm HF}_\alpha(\bs{k})$ with quasiparticle creation operators $\psi^\dagger_{\bs{k},\alpha}=U^{\rm HF}_{\alpha a}c_{\bs{k} a}^\dagger$, where indices $\alpha=0,1,2\ldots$ are ordered by increasing energy at every $\bs{k}$.

To study self-doping of electron crystals, we minimize the HF ground-state energy as a function of the density mismatch $\delta n = n_e - n_{\rm u.c.}$ over the range $|\delta n| \le 0.25\times 10^{12}\,\mathrm{cm}^{-2}$. A conventional commensurate insulating WC has $\delta n = 0$, while $\delta n^* \neq 0$ signals a self-doped, incommensurate metallic WC. Figure~\ref{fig:1_HF_self-doping}(b) presents the evolution of the optimal self-doping as a function of electronic density for a fixed displacement field of $u_D=65$~meV.  At small electronic density, h-MWCs with a small simply connected hole Fermi surface (FS) surrounding the BZ center are favored (Fig.~\ref{fig:1_HF_self-doping}(c)). As the electronic density increases, band 1 crosses the chemical potential around the $\kappa^+$ high-symmetry point. This leads to an intermediate regime where a hole FS in band 0 coexists with an electron FS in band 1, as illustrated in Fig.~\ref{fig:1_HF_self-doping}(d). Eventually, the lowest HF band falls entirely below the chemical potential, and an e-MWC with a single electron FS remains at larger densities (Fig.~\ref{fig:1_HF_self-doping}(b) and (e)). This evolution appears broadly consistent with previous theoretical investigations~\cite{dongCrystalsCaughtDoping2026, fengSelfdopedCrystalPreempted2026} and with experimental reports of a hole-doped metallic crystal at low densities near the insulating WC regime in R(4,5,6)G~\cite{hanEvidenceMetallicWigner2026}, as well as claims of electron-doped metallic WC at higher densities in R5G~\cite{zhouCompetingOrdersDriven2026a} and R6G~\cite{nguyenCoexistingChargeDensity2026} (see Fig.~\ref{fig:1_HF_self-doping}(a)). 

\begin{figure*}[t]
    \centering
    \includegraphics[width=1.0\textwidth]{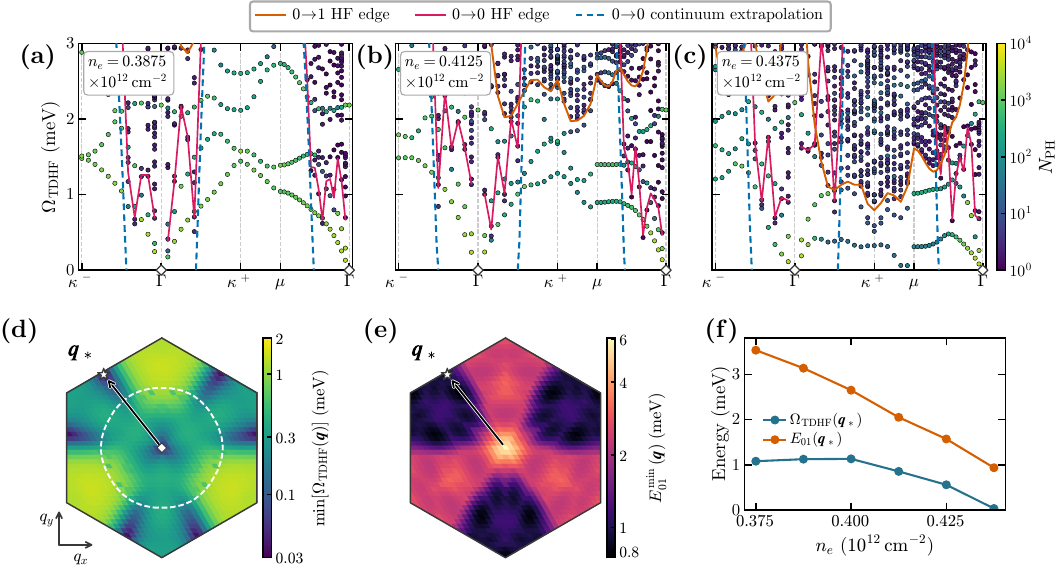}\vspace{-2mm}
    \caption{
    \textbf{Density evolution of collective-mode softening.} 
    (a--c) TDHF spectra along high-symmetry line at $n_e=(0.3875,0.4125,0.4375)\times10^{12}\,\mathrm{cm}^{-2}$, colored by particle--hole participation number $N_{\rm PH}$ (see Eq.~\eqref{eq:tdhf-participation}). Lines show the HF continuum edges. (d,e) Lowest TDHF energy and $0\to1$ continuum threshold across the Brillouin zone at $n_e=0.4375\times10^{12}\,\mathrm{cm}^{-2}$. The dashed contour bounds the gapless $0\to0$ continuum region, and stars mark $\boldsymbol{q}_*$. (f) Density dependence of both energies at this reduced momentum.
    }
    \label{fig:2_collective_mode_and_softening}
\end{figure*}

%%%%%%%%%%%%%%%%%%%%%%%%%%%%%%%%%%%%%%
% Collective modes and instabilities %
%%%%%%%%%%%%%%%%%%%%%%%%%%%%%%%%%%%%%%
\textit{Collective modes and instabilities}.---Having established the HF MWC states and their evolution as a function of electronic density, we next examine their collective excitations using the time-dependent Hartree--Fock (TDHF) approximation, also known as the generalized random-phase approximation~\cite{thouless1961vibrational, thouless1962time, rowe1968equations, cote1991collective, vasiliou2024electrons}. Linearizing the dynamics about each stationary HF solution yields the following generalized eigenvalue problem for the collective modes,
%~\footnote{From a diagrammatic point of view, this is obtained by solving the Bethe-Salpeter equation for the particle-hole propagator using direct (Hartree) and exchange (Fock) channels for the particle-hole irreducible vertex (i.e., summing both bubble and ladder diagrams). It is equivalent to applying the time-dependent variational principle~\cite{kramer1981geometry, dong2025phonons} and is a conserving approximation in the sense of Baym and Kadanoff~\cite{baym1961conservation}.}
\begin{equation}
 \mathcal H_{\bs q}
 \begin{pmatrix}X_{\bs q\nu}\\Y_{\bs q \nu}\end{pmatrix}
 =\Omega_{\nu\bs q}
 \begin{pmatrix}\mathds{1}&0\\0&-\mathds{1}\end{pmatrix}
 \begin{pmatrix}X_{\bs q \nu}\\Y_{\bs q \nu}\end{pmatrix}.
 \label{eq:tdhf-spectrum}
\end{equation}
Here $\bs q$ lies in the crystal's first Brillouin zone, $\Omega_{\bs q \nu}$ is the excitation energy, and $X_{\bs q\nu}$ and $Y_{\bs q \nu}$ are the forward and backward particle--hole amplitudes of mode $\nu$. The fluctuation Hessian $\mathcal H_{\bs q}$ contains the HF particle-hole energy differences and the residual Hartree and exchange interactions within the seven-band space. For real modes with positive norms, the associated charge-neutral excitations have quasi-boson creation operators $Q_{\nu\bs q}^{\dagger} =\sum_{i,\bs k}X^i_{\nu\bs q}(\bs k)b_{i,\bs q}^{\dagger}(\bs k) - \sum_{i,\bs k}Y^i_{\nu\bs q}(\bs k)b_{i,-\bs q}(\bs k)$, where $i=(\phi_p,\phi_h)$ labels pairs of particle $(\phi_p)$ and hole $(\phi_h)$ and $b_{i,\bs q}^{\dagger}(\bs k) =\psi_{\phi_p,\bs k+\bs q}^{\dagger}\psi_{\phi_h,\bs k}$ is a particle-hole creation operator. Each sum runs over the occupied-to-empty transitions at its indicated transfer momentum. Explicit matrix elements and numerical details are given in the Supplemental Material~\cite{Note2}. 

Figures~\ref{fig:2_collective_mode_and_softening}(a)-(c) show the excitation spectra at $u_D=65\,\mathrm{meV}$ in the h-MWC regime for $n_e=0.3875,0.4125,$ and $0.4375\times10^{12}\,\mathrm{cm}^{-2}$. The TDHF response contains collective modes, such as two gapless Goldstone phonon modes arising from the spontaneous breaking of continuous translational symmetry, together with continuum-like particle-hole excitations that correspond to moving a filled state in the HF band structure to an empty one. To qualitatively distinguish between these two types of modes, we color each mode by its particle--hole participation number,
\begin{equation}
    N_{\rm PH}^{\nu\bs q} =\frac{\left(\sum_r|X_r|^2+\sum_s|Y_s|^2\right)^2}{\sum_r|X_r|^4+\sum_s|Y_s|^4}, \label{eq:tdhf-participation}
\end{equation}
where $r$ and $s$ enumerate the complete forward and backward configurations, including their band and momentum labels, and the mode labels on $X$ and $Y$ are suppressed. The particle--hole participation number is unity for a single-component excitation and equals $m$ when the weight is evenly distributed among $m$ components.

We observe bands of large $N_{\rm PH} $ values, corresponding to the collective modes, as well as large groups of modes with $N_{\rm PH}$ near unity, corresponding to the continuum of non-interacting particle-hole excitations above the HF ground state. We confirm this interpretation by additionally plotting in Fig.~\ref{fig:2_collective_mode_and_softening}(a)-(c) the lower edges of the noninteracting HF particle--hole continua, $E_{\alpha\beta}^{\min}(\bs q) =\min_{\bs k} \left[\varepsilon_\beta^{\rm HF}(\bs k+\bs q) -\varepsilon_\alpha^{\rm HF}(\bs k)\right]$, where the ($\alpha,\bs{k}$) HF state is occupied and ($\beta, \bs{k} + \bs{q})$ is empty. The $0\to0$ and $0\to1$ thresholds are shown by solid curves, and the dashed curve estimates the intraband continuum edge that would be recovered in the thermodynamic limit.

Studying the evolution of these two types of modes as the density increases, we see the $0\to1$ continuum gradually lowering in energy around the $\kappa^+$ point, anticipating the nucleation of the electron FS in the second band. As the $0\to1$ continuum drops, it hybridizes with the phonon collective modes (as evidenced by the reduced value of $N_{\rm PH}$ in those bands) and pushes them to lower energies. At $n_e=0.4375\times10^{12}$~cm$^{- 2}$, the lowest collective mode becomes critically soft along the $\kappa^+\to\mu$ line and loses most of its collective character. This behavior suggests an effective level repulsion between the collective mode and the $0\to1$ continuum, reminiscent of phenomena observed in other systems~\cite{pitaevskii1959properties, gaveau1995limited, verresen2019avoided, wang2024interaction}. In Fig.~\ref{fig:2_collective_mode_and_softening}(d-f), we plot the minimum energy of the TDHF modes and the minimum of the non-interacting $0\to1$ continuum as functions of momentum. This data shows that the phonon softens at $\bs{q}_*$ around the bottom of the $0\to1$ continuum, confirming that the mode softening correlates with the evolution of the continuum. For larger carrier densities, above $n_e\ge 0.45\times 10^{12}$~cm$^{-2}$, the collective mode spectrum begins to exhibit complex finite-momentum TDHF frequencies simultaneously with the nucleation of the electron Fermi pocket, indicating dynamical instabilities.

These results raise a few important points worth careful consideration when asking whether collective modes of the crystal can mediate pairing. In addition to the long-wavelength phonons, the finite-momentum soft sector may make an important contribution to the retarded pairing interaction, especially where it is very soft just before the electron Fermi surface begins to form. This finite-momentum soft sector can provide a strong pairing contribution beyond the long-wavelength phonons. In the same regime, pairing between bands 0 and 1 may be particularly relevant, since the minimum of the first band lies only just above the chemical potential. Furthermore, the momentum transfer $\bs{q}_*$ of the soft mode connects states on the Fermi surface to the low-energy portion of this first band. As such, limiting a pairing analysis to the Fermi surface, the lowest HF band, or the interaction mediated by long-wavelength phonons near the zone center may miss crucial physics.

%%%%%%%%%%%%%%%%%%%%%%%%%%%%%%%%%%%%%%%%%%%%%%%
% Effective interaction and superconductivity %
%%%%%%%%%%%%%%%%%%%%%%%%%%%%%%%%%%%%%%%%%%%%%%%
\textit{Effective interaction and superconductivity}.---We next investigate whether the collective fluctuations of the electron crystal can mediate superconducting pairing. The self-consistent HF solution provides the reconstructed electronic bands and wave functions, while TDHF determines the dynamics of fluctuations about this reference state and their coupling to the electrons~\cite{bickers1989conserving, schrieffer1989dynamic, chubukov1992renormalized, vasiliou2024electrons}. At Gaussian order, these ingredients define an electron–boson action. Integrating out the bosonic fields yields a frequency-dependent electronic interaction that contains both the instantaneous Coulomb contribution and a retarded contribution inherited from the TDHF response, $\mathcal{W}_{\alpha\beta}(\bs{q},i\omega_n) $~\cite{Note2}. We retain the full momentum and Matsubara-frequency dependence of the effective interaction, including the effects of collective fluctuations beyond the modes identified as phonon-like. 

\begin{figure}
    \centering
    \includegraphics[width=1.0\linewidth]{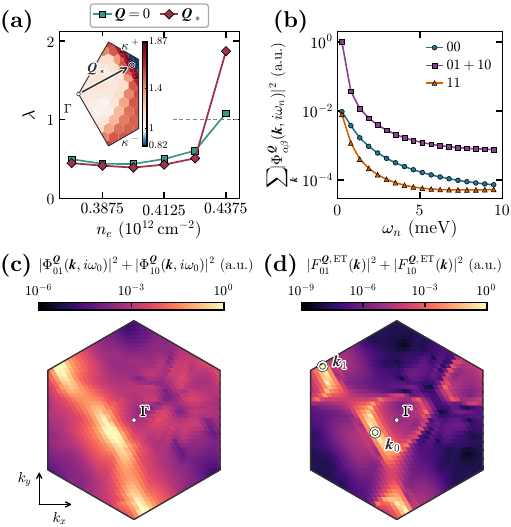}
    \caption{
    \textbf{Emergence and structure of finite-momentum interband pairing.}
    (a) Leading nearly real pairing eigenvalues at $\boldsymbol{Q}=0$ and the common finite momentum $\boldsymbol{Q}_*$ versus density. The inset shows the momentum dependence over one-third of the first Brillouin zone at $n_e=0.4375\times10^{12}\,\mathrm{cm}^{-2}$. The arrow identifies momentum $\boldsymbol{Q}_*$ with the largest pairing eigenvalue. The dashed line marks $\lambda=1$. (b) Band-resolved frequency dependence of the gap intensity for the leading mode at $\boldsymbol{Q}_*$. (c,d) Anomalous self-energy at the lowest positive Matsubara frequency and equal-time pairing-response intensities. Markers in (d) identify the dominant partner states $\boldsymbol{k}_0$ and $\boldsymbol{k}_1$. Results use fixed HF propagators at $T=1\,\mathrm{K}$.
    }
    \label{fig:3_SC}
\end{figure}

With this effective interaction between the HF single-particle states, we study pairing within the occupied spin–valley flavor, retaining HF bands $\alpha,\beta=0,1$ throughout the crystal Brillouin zone. For each total pair momentum $\bs Q$, we solve the linearized Eliashberg equation, formulated as an eigenvalue problem in the particle-particle channel~\cite{eliashberg1960interactions, bickers1989conserving, scalapino2012common, takimoto2004strong, allen1983theory, carbotte1990properties},
\begin{equation}\label{eq:linearized_eliashberg}
\begin{aligned}
&\lambda_{\bs Q}\Phi^{\bs Q}_{\alpha\beta}(\bs k,i\omega_n) \\
& =
\frac{k_BT}{2N_k} \sum_{\bs k',i\omega_m,\mu\nu} \Gamma^{A,\bs Q}_{\alpha\beta;\mu\nu}(\bs k, i\omega_n;\bs k', i\omega_m) F^{\bs Q}_{\mu\nu}(\bs k',i\omega_m),
\end{aligned}
\end{equation}
where $\Phi$ is the anomalous self-energy, $F$ is the induced anomalous response
\begin{equation}\label{eq:induced_anomalous_response}
\begin{aligned}
&F^{\bs Q}_{\alpha\beta}(\bs k,i\omega_n)\\
&\quad=
G^{\mathrm{HF}}_\alpha(\bs k,i\omega_n)\,
\Phi^{\bs Q}_{\alpha\beta}(\bs k,i\omega_n)\,
G^{\mathrm{HF}}_\beta(\bs Q-\bs k,-i\omega_n),
\end{aligned}
\end{equation}
and $\Gamma^A$ is the antisymmetrized pairing kernel that includes the effect of the instantaneous Coulomb contribution and retarded interaction mediated by TDHF fluctuations. Fermionic antisymmetry exchanges the band indices, partner momenta, and Matsubara frequencies together, which imposes $\Phi_{a b}^{\mathbf{Q}}\left(\mathbf{k}, i \omega_n\right)=-\Phi_{b a}^{\mathbf{Q}}\left(\mathbf{Q}-\mathbf{k},-i \omega_n\right)$. We keep the normal propagators fixed at their HF values, $G_\alpha^{\mathrm{HF}}=[i\omega_n-\xi_\alpha(\bs k)]^{-1}$ with $\xi_\alpha(\bs k)=\varepsilon_\alpha^{\rm HF}(\bs{k}) - \mu$, rather than updating them through a normal self-energy calculation, i.e., a fully self-consistent Eliashberg treatment~\cite{eliashberg1960interactions, carbotte1990properties}. A real eigenvalue that reaches $\lambda_{\mathbf Q}=1$ determines the onset of a superconducting instability within the linearized Eliashberg approximation~\footnote{However, we emphasize that the corresponding temperature need not coincide with the actual superconducting transition temperature, as fluctuations beyond this approximation, particularly phase fluctuations in two dimensions, can suppress the establishment of superconducting phase coherence~\cite{kosterlitz1973ordering, scalapino1992superfluid, emery1995importance}.}.

Figure~\ref{fig:3_SC}(a) shows the evolution of the largest pairing eigenvalue at $T=1\,\mathrm K$ for carrier densities spanning the h-MWC regime (see Fig.~\ref{fig:1_HF_self-doping}(b)). We find that the largest pairing eigenvalue occurs at the largest carrier density studied, $n_e=0.4375\times10^{12}~\mathrm{cm}^{-2}$, for a finite-momentum $\bs{Q}_*$ depicted in the inset of Fig.~\ref{fig:3_SC}(a). We plot the largest pairing eigenvalues at both $\bs{Q}=0$ and $\bs{Q}=\bs{Q}_*$ in Fig.~\ref{fig:3_SC}(a), both of which remain below unity for low carrier densities. However, a pronounced enhancement occurs in the regime of strong collective-mode softening at $n_e=0.4375\times10^{12}\,\mathrm{cm}^{-2}$, with finite-momentum pairing obtaining a dominant eigenvalue of $\lambda_{\bs{Q}_*}\simeq1.87$. These results identify a strong finite-momentum pairing tendency in the fixed-HF calculation close to the onset of the collective instability.

As shown in Fig.~\ref{fig:3_SC}(b), the corresponding eigenmode is predominantly interband. At the lowest Matsubara frequency, the $01$ and $10$ components account for approximately $98.2\%$ of the squared norm of the gap, demonstrating that band 1 participates strongly in the pairing even before it develops an electron Fermi surface. A calculation restricted to band 0 would therefore miss the dominant pairing channel. To visualize the momentum structure of the pairing eigenmode, we present in Fig.~\ref{fig:3_SC}(c,d) the lowest Matsubara frequency component of the interband anomalous self-energy and the interband part of the equal-time anomalous response $F_{a b}^{\mathrm{Q},\mathrm{ET}}(\mathbf{k})=k_B T \sum_{n=-\infty}^{\infty} F_{a b}^{\mathrm{Q}}\left(\mathbf{k}, i \omega_n\right)$. Both quantities are sharply concentrated in momentum, especially the anomalous response, with $99.93\%$ of its weight interband. The maxima of the equal-time response occur at $\bs k_0$ and $\bs k_1$, related by $\bs k_0+\bs k_1=\bs{Q}_*$ modulo a reciprocal lattice vector, representing the two band-resolved partners of the same finite-momentum pair. 

\begin{figure}[t]
    \centering
    \includegraphics[width=1.0\linewidth]{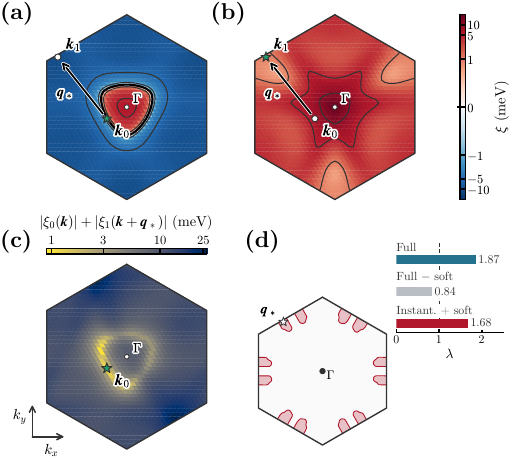}
    \caption{
    \textbf{Soft-mode contribution to finite-momentum pairing.}
    (a,b) HF dispersions of bands 0 and 1, marking the dominant pairing partners $\boldsymbol{k}_0$ and $\boldsymbol{k}_1$ and their separation $\boldsymbol{q}_*$. (c) Energy cost $|\xi_0(\boldsymbol{k})|+|\xi_1(\boldsymbol{k}+\boldsymbol{q}_*)|$, illustrating the energetic compatibility of states connected by this transfer. (d) The selected momentum regions corresponding to the soft mode. The inset on the top right shows independently optimized pairing eigenvalues for the full interaction, with the soft sector removed, and with only the instantaneous interaction plus the soft sector retained. All panels concern the leading state at $n_e=0.4375\times10^{12}\,\mathrm{cm}^{-2}$ in Fig.~\ref{fig:3_SC}.
    }
    \label{fig:4_SC_from_critical_mode}
\end{figure}

%%%%%%%%%%%%%%%%%%%%%%%%%%%%%%%%%%%%%%%
% Pairing from crystal critical modes %
%%%%%%%%%%%%%%%%%%%%%%%%%%%%%%%%%%%%%%%
\textit{Pairing from crystal critical modes}.---The unusual structure of the gap we observe (i.e., finite-momentum, interband, with a strongly localized response) motivates a closer examination of why collective fluctuations may favor such unconventional pairing. First, we note that the two momenta involved in the pairing response, $\bs{k}_0$ in band 0 (Fig.~\ref{fig:4_SC_from_critical_mode}(a)) and $\bs{k}_1$ in band 1 (Fig.~\ref{fig:4_SC_from_critical_mode}(b)), are separated by the momentum of the soft collective mode we identified in Fig.~\ref{fig:2_collective_mode_and_softening}: $\bs{k}_1-\bs{k}_0=\bs{q}_*$. This key observation strongly implies that pairing is driven by the soft mode.

To affirm the central role that the soft mode plays in pairing, we repeat the pairing analysis, isolating its contribution. We first define the finite-momentum soft sector as the locus of the lowest positive-norm TDHF modes with energies below \(0.1\,\mathrm{meV}\) in the three symmetry-related pockets disconnected from the \(\Gamma\)-centered low-energy region, as depicted in Fig.~\ref{fig:4_SC_from_critical_mode}(d). To assess their contribution to pairing, we remove their pole contributions to the retarded interaction and solve the linearized pairing equation~\eqref{eq:linearized_eliashberg} again at the same temperature and pair momentum, keeping the HF propagators and instantaneous interaction fixed. As shown in Fig.~\ref{fig:4_SC_from_critical_mode}(d) the leading pairing eigenvalue at $\bs{Q}_*$ drops below 1 to 0.84. In contrast, if we redo the analysis but instead include only the contribution from these soft modes and the instantaneous Coulomb repulsion, we still observe a pairing instability with a substantial pairing eigenvalue of 1.68 (see inset of Fig.~\ref{fig:4_SC_from_critical_mode}(d)). This analysis provides unambiguous evidence that the strongest pairing we observe is mediated by these soft modes.

The soft mode mediating the pairing implies a specific separation between the pair momenta $\bs{k}_0$ and $\bs{k}_1$. To understand how it further leads to a finite-$\bs{Q}$ instability, we plot in Figs.~\ref{fig:4_SC_from_critical_mode}(a) and (b) the HF dispersion of bands 0 and 1. We see that $\bs{k}_0$ sits on the Fermi surface in band 0, whereas $\bs{k}_1$ is located around the minimum of band 1, where the incipient electron FS will form.  The soft mode fixes the relative momentum of the paired electrons, but not their total momentum $\bs{Q}$. Exchanging a soft mode moves one electron from $(\bs{k}_0,0)$ to $(\bs{k}_1,1)$ and its partner from $(\bs{k}_1,1)$ back to $(\bs{k}_0,0)$, so the pair is scattered back onto itself provided its two electrons are separated by $\bs{k}_1-\bs{k}_0=\bs{q}_*$. Such a pair carries total momentum $\bs{Q}=\bs{k}_0+\bs{k}_1=2\bs{k}_0+\bs{q}_*$, and the remaining question is which $\bs{k}_0$ (i.e., $\bs{Q}$) is selected. Heuristically, pairing is strongest when both partners lie close to the chemical potential, that is, when the pair energy $|\xi_0(\bs{k})|+|\xi_1(\bs{k}+\bs{q}_*)|$ is smallest. Figures~\ref{fig:4_SC_from_critical_mode}(a) and (b) show that band 0 disperses steeply across the hole FS, whereas band 1 has a shallow minimum just above the chemical potential, where the electron pocket will later form. The pair energy is therefore minimized by placing $\bs{k}_0$ on the hole FS such that $\bs{k}_0+\bs{q}_*$ falls at the bottom of band 1, as confirmed in Fig.~\ref{fig:4_SC_from_critical_mode}(c). This naturally selects a finite pair momentum $\bs{Q}$. The same argument explains why the pairing is interband. As $\bs{q}_*$ lies outside the region of Fig.~\ref{fig:2_collective_mode_and_softening}(d) where the intraband continuum is gapless, it cannot connect two points of the hole FS, and band 0 at $\bs{k}_1$ lies far from the chemical potential. Finally, we note that this picture is fully consistent with the observation in Fig.~\ref{fig:3_SC}(a) that the pairing tendency is only strongly enhanced at large densities close to the collective instability where the mode is the softest.

%%%%%%%%%%%%%%
% Discussion %
%%%%%%%%%%%%%%
\textit{Discussion}.---We have studied possible pairing instabilities of metallic crystals in R4G that could be mediated by collective electronic modes of the crystal. We find a finite-$\bs{Q}$ interband pairing instability that results in a state with broken time-reversal symmetry and coexisting charge ordering and modulated superconductivity---an electronic ``super-supersolid''~\footnote{F.D. acknowledges Zhaoyu Han for proposing the name super-supersolid. F.D. also acknowledges Pavel Nosov for instead proposing supersolidsolid or solid supersolid}. The additional ``super'' reflects the finite-$\bs{Q}$ pairing of the SC. In particular, a triple-$\bs{Q}$ condensate (PDW) would have some additional charge (commensurate or incommensurate) modulation. Thus leading to a ``superlattice'' of charge modulation atop the WC. We further identify that such pairing is mediated by a soft mode associated with the onset of an instability that appears when the electron crystal goes from hole- to electron-doped and develops an electron Fermi surface around $\kappa^+$. This situation is remarkable, representing a far more sophisticated phenomenon than the conventional phonon-mediated pairing of electrons on the FS that one may have naively anticipated. Furthermore, this result reproduces the overall trend observed experimentally, that of a h-MWC at low carrier densities that transitions to a superconductor and eventually to an e-MWC at larger densities (see Fig.~\ref{fig:1_HF_self-doping}(a)). The soft-mode-mediated pairing also provides important insight into why superconductivity is observed only in a restricted region of the phase diagram, rather than throughout the whole region where MWCs are present.

Our findings suggest several possible experimental signatures of the putative state. The most obvious one is the presence of Wigner crystalline order in the superconducting state, which, although challenging to image directly in dual-gated devices, could be accessed through quantum twisting microscopy or other indirect probes~\cite{Clemens}. The coexistence of superconductivity and charge order could also manifest indirectly in other probes, such as nonlinear transport, wherein crystal depinning and superconducting depairing may occur at distinct current scales, leading to distinct transport regimes. Althought we have not determined a physical superconducting critical temperature, an interesting possibility is that the critical temperature may be limited not by pairing or superfluid stiffness, but rather by the formation of the crystal itself. In such a case, the superconducting transition would occur simultaneously with the crystallization transition and would be first order. This appears reasonable, given that the melting point of the metallic Wigner crystal has been estimated to be on the order of 250~mK in Ref.~\cite{dongCrystalsCaughtDoping2026}, which roughly matches the critical temperature of the superconductors observed experimentally. Finally, the finite-momentum nature of the pairing could lead to distinct experimental signatures in Josephson interferometry~\cite{Chen2018} or spatially resolved superconducting spectroscopy~\cite{Gu2023}. The threefold rotational symmetry of the electron crystal implies three degenerate pairing wavevectors, raising the possibility of either a nematic, single-$\bs{Q}$ Fulde--Ferrell-like state or a $C_3$-symmetric, triple-$\bs{Q}$ pair-density wave. These possibilities could be distinguished by their rotational symmetry and spatial pairing structure using angle-dependent superconducting transport~\cite{Pan2016} or scanning Josephson tunneling microscopy~\cite{Hamidian2016}.

We note that there are already some hints of crystalline order being entwined with superconductivity in rhombohedral graphene. In Ref.~\cite{nguyenCoexistingChargeDensity2026}, a superconductor stabilized only in the presence of a perpendicular magnetic field was observed in low-density rhombohedral hexalayer graphene. Upon exiting the superconductor by warming or lowering the magnetic field, hysteresis is observed, marking a first-order transition. A natural explanation for the surprising first-order transition into the superconductor is that the underlying crystalline order, probably stabilized by the magnetic field, is essential to superconductivity, so melting the crystal also destroys superconductivity. In a different experiment~\cite{kalantreFermiologyCandidateChiral2026a}, multitone quantum oscillations were observed above a candidate chiral superconductor in rhombohedral tetralayer graphene. The apparent deficit of itinerant carriers in an annular Fermi surface model of these observations motivates the picture in which some electrons are localized in a coexisting crystal~\cite{guerciQuantumOscillationsHole2026}. More experiments are needed to test this scenario.

Several extensions and refinements of our work remain important. First, while our linearized Eliashberg analysis identifies the leading pairing instability, determining the superconducting transition temperature requires tracking its temperature dependence and accounting for the superconducting phase stiffness~\cite{kosterlitz1973ordering, scalapino1992superfluid, emery1995importance}.  Our analysis further neglects the electronic self-energy generated by collective modes and the feedback of the associated renormalized propagator on the collective-mode self-energy. It would thus be interesting to test the robustness of our conclusion as one moves towards a full self-consistent Migdal-Eliashberg treatment of the pairing problem~\cite{eliashberg1960interactions, carbotte1990properties}. An assessment of the strength of vertex corrections would also be desirable, particularly given the soft electronic modes responsible for pairing. Furthermore, determining whether the superconducting state selects a single pairing momentum or a coherent superposition of symmetry-related momenta would be particularly relevant for making specific experimental predictions, but it requires an analysis of the nonlinear superconducting free energy, through, for example, a Ginzburg–Landau expansion~\cite{fulde1964superconductivity, larkin1964nonuniform, larkin1965nonuniform, shimahara1998structure, agterberg2020physics, bowers2002crystallography}. Similarly, determining whether the zero-temperature superconducting state is fully gapped, and if so its topological characterization, requires going beyond the linear stability analysis. Finally, the softening of the collective mode ultimately signals an instability of the underlying electron crystal, beyond which our present approach is no longer applicable. Treating crystallization and superconductivity on equal footing could reveal whether superconducting order preempts this instability, potentially stabilizing superconductivity in an even wider region of the phase diagram~\cite{balseiro1979superconductivity, bilbro1976theoretical, fernandes2010competing}.

Our results suggest a general mechanism by which interactions promote superconductivity, in which pairing is driven by interband soft modes. The relevant ingredients are a Fermi surface and a weakly dispersing unoccupied band approaching the chemical potential. Interband particle–hole excitations connecting them can drive a finite-wave-vector instability of the crystal. Before that instability develops, the soft mode can mediate a strong pairing interaction to pair carriers across the two bands. Incipient flat bands and proximity to electronic-crystal instabilities may therefore provide favorable conditions for unconventional superconductivity. More broadly, our findings indicate that the electronic correlations responsible for crystallization can also provide the mechanism for superconducting pairing. This raises the tantalizing prospect that an electronic super-supersolid has already been realized in rhombohedral graphene, but has so far escaped experimental identification. Our findings thus motivate the search for coexisting charge order and superconductivity in these systems.

\textit{Note added}.---During the preparation of this manuscript, Ref.~\cite{Kim2026} appeared, independently identifying the same collective-mode instability in metallic Wigner crystals. While superconductivity is not addressed in that work, the results are consistent with ours in their common regime of applicability. We were also made aware of concurrent work~\cite{Zhang2026} investigating superconductivity in metallic Wigner crystals, which should have partial overlap with ours and will appear in a forthcoming preprint.

\begin{acknowledgments}
\textit{Acknowledgments}.---We thank Pavel Nosov, Zhaoyu Han, and Clemens Kuhlenkamp for stimulating and helpful discussions. We further thank Zhaoyu Han and Zhihuan Dong for informing us of their involvement in concurrent work and coordinating submission. Our computations were run on the FASRC Cannon cluster supported by the FAS Division of Science Research Computing Group at Harvard University. A.V. and F.D. acknowledge funding from NSF DMR-2220703 and the Simons Collaboration on Ultra-Quantum Matter, which is funded by the Simons Foundation (651440, A.V.).  M.R.H. and Y.B.K. acknowledge support from the Natural Science and Engineering Council of Canada (NSERC) Grant No. RGPIN-2023-03296 and the Center for Quantum Materials at the University of Toronto. The authors acknowledge the use of AI tools ChatGPT 6 Astra and Claude Opus 5.5 for assistance with coding and producing the figures. All AI-assisted results were independently verified by the authors, who take full responsibility for the content of this work.
\end{acknowledgments}

\bibliography{ref}

% \clearpage
% %%%%%%%%%%%%%%%%%%%%%%%%%%%%%%%%%%%%%%%%%%%%%%%%%%
% %                  End Matter                    %
% %%%%%%%%%%%%%%%%%%%%%%%%%%%%%%%%%%%%%%%%%%%%%%%%%%
% \onecolumngrid
% \section*{End Matter}
% \twocolumngrid

\end{document}

\renewcommand{\thesection}{S.\Roman{section}}
\renewcommand{\thesubsection}{\Alph{subsection}}
\renewcommand{\theequation}{S\arabic{equation}}
\renewcommand{\thefigure}{S\arabic{figure}}
\renewcommand{\thetable}{S\arabic{table}}

\title{%
Supplemental Material for: \linebreak
``Superconductivity in the Metallic Wigner Crystals of Rhombohedral Graphene:\\
Possible Realization of an Electronic Supersolid\,''
}
\author{F\'elix Desrochers\,\orcidlink{0000-0003-1211-901X}}
%\email{felix\_desrochers@fas.harvard.edu}
\affiliation{Department of Physics, Harvard University, Cambridge, MA 02138, United States}
\author{Mark R. Hirsbrunner\,\orcidlink{0000-0001-8115-6098}}
\affiliation{Department of Physics, University of Toronto, Toronto, Ontario M5S 1A7, Canada}
\author{Yong Baek Kim\,\orcidlink{0000-0003-0473-8785}}
\affiliation{Department of Physics, University of Toronto, Toronto, Ontario M5S 1A7, Canada}
\author{Ashvin Vishwanath\,\orcidlink{0000-0002-6306-2263}}
%\email{avishwanath@g.harvard.edu}
\affiliation{Department of Physics, Harvard University, Cambridge, MA 02138, United States}
\date{\today}
\maketitle
\vspace{-\baselineskip}

% A compact contents list: main sections only.
\setcounter{tocdepth}{1}
\begingroup
\small
% REVTeX does not use tocdepth to suppress subsection entries.
\makeatletter
\let\l@subsection\@gobbletwo
\let\l@subsubsection\@gobbletwo
\def\l@f@section{\addpenalty{\@secpenalty}\addvspace{0.15em}}
\makeatother
\hypersetup{linkcolor=black}
\tableofcontents
\endgroup
\vspace{0.5\baselineskip}

%==============================================================================
\section{Continuum model}
%==============================================================================

\subsection{Single-particle Hamiltonian}
We consider rhombohedral tetralayer graphene in a single spin and valley sector, assuming complete flavor polarization. We use the tight-binding description of Refs.~\cite{dongCrystalsCaughtDoping2026,auerbach2025isospin}. In the layer and sublattice basis $(A_1,B_1,A_2,B_2,A_3,B_3,A_4,B_4)$, with $A$ and $B$ denoting the graphene sublattice and the subscript denoting the layer index, the single-particle Hamiltonian is
\begin{equation}
 h_{\rm R4G}(\bs p)=
 \begin{pmatrix}
 D_1&T&W&0\\ T^\dagger&D_2&T&W\\
 W^\dagger&T^\dagger&D_3&T\\0&W^\dagger&T^\dagger&D_4
 \end{pmatrix},
\end{equation}
where
\begin{equation}
 D_\ell=\begin{pmatrix}u_\ell+d_{\ell A}&-t_0f_{\bs p}\\
                  -t_0f_{\bs p}^*&u_\ell+d_{\ell B}\end{pmatrix}, \quad
 T=\begin{pmatrix}t_4f_{\bs p}&t_3f_{\bs p}^*\\t_1&t_4f_{\bs p}\end{pmatrix},\quad
 W=\begin{pmatrix}0&t_2/2\\0&0\end{pmatrix}.
\end{equation}
The diagonal blocks describe intralayer hopping and onsite potentials, whereas the off-diagonal blocks couple adjacent and next-nearest layers. The momentum $\bs p$ is measured relative to the selected valley in crystal coordinates. We retain the full graphene structure factor,
\begin{equation}
 f_{\bs p}
 =\sum_{j=0}^{2}
 e^{i(\bs K+R_{-\pi/6}\bs p)\cdot\bs\delta_j},
\end{equation}
where $\bs K=(4\pi/3a,0)$, $\bs\delta_j=(a/\sqrt{3})R_{2\pi j/3}(0,1)$ are the nearest-neighbor vectors, and $R_\theta$ denotes a counterclockwise rotation by $\theta$. The rotation $R_{-\pi/6}$ relates the crystal and graphene axes, and $a=0.246\,\mathrm{nm}$ is the graphene lattice constant.

The displacement field produces the layer potential $u_\ell=u_D(\ell-5/2)$, where $u_D=65\,\mathrm{meV}$ is the potential difference between adjacent layers. The onsite offsets are $d_{1A}=d_{4B}=\delta_2$, $d_{1B}=d_{4A}=\delta+\delta_2$, and $d_{2A}=d_{2B}=d_{3A}=d_{3B}=\delta-\delta_2$, with $\delta=10.5\,\mathrm{meV}$ and $\delta_2=2\,\mathrm{meV}$. The hopping amplitudes are $t_0=3100\,\mathrm{meV}$, $t_1=380\,\mathrm{meV}$, $t_2=-15\,\mathrm{meV}$, $t_3=290\,\mathrm{meV}$, and $t_4=141\,\mathrm{meV}$.

\subsection{Projected basis and screened interaction}

We allow for triangular crystalline order in the absence of an external periodic potential. The crystal is generated by
\begin{equation}
 \bs a_1=L\left(\tfrac{\sqrt3}{2},\tfrac12\right),\qquad \bs a_2=L(0,1),
\end{equation}
so the unit cell has area $A_{\rm uc}=\sqrt3L^2/2$ and one electron per cell corresponds to the density $1/A_{\rm uc}$. The reciprocal vectors are $\bs b_1=(4\pi/\sqrt3L)(1,0)$ and $\bs b_2=(4\pi/\sqrt3L)(-1/2,\sqrt3/2)$. We sample the crystal Brillouin zone (BZ) with a $36\times36$ mesh given by
\begin{equation}
 \bs k=\frac{m_1\bs b_1+m_2\bs b_2}{36},\qquad
 m_1,m_2\in\{-18,\ldots,17\},
\end{equation}
so the sample contains $N=36^2$ crystal cells. Reduced momenta are defined modulo the reciprocal lattice.

We project onto the lowest conduction band of $h_{\rm R4G}$. Folding this band into the crystal BZ gives, at each $\bs k$, the states $|\bs k+\bs G,4\rangle_{\rm R4G}$, where 4 labels the first conduction band in the zero-based ordering of the eight bands. Among the 19 reciprocal vectors with $|\bs G|\le2|\bs b_1|$ ($\bs G=0$ and the first three shells), we keep the seven states lowest in energy,
\begin{equation}
 \operatorname{span}\{|\bs k,a\rangle\}_{a=0}^{6}
 =\underset{\text{seven lowest energies}}{\operatorname{span}}
 \{|\bs k+\bs G,4\rangle_{\rm R4G}:|\bs G|\le2|\bs b_1|\}.
 \label{eq:projected-basis}
\end{equation}
These are full continuum Bloch states, including their layer and sublattice structure, and the retained subspace is held fixed during the Hartree--Fock (HF) iteration.
% At cutoff degeneracies, the original saved choice of retained states is kept.

For convenience, we use bold numerals as composite labels, $\mathbf 1\equiv(\bs k_1,a_1)$, with $\langle\mathbf1|\mathbf2\rangle=\delta_{\mathbf1\mathbf2}$. in this basis, the kinetic energy $h^0_{\mathbf1\mathbf2}=\langle\mathbf1|\widehat h_{\rm R4G}|\mathbf2\rangle$ is diagonal in $\bs k$, and the density form factor is
\begin{equation}
 \mathcal F_{\mathbf1\mathbf3}(\bs q)=\langle\mathbf1|e^{i\bs q\cdot\hat{\bs r}}|\mathbf3\rangle,
 \qquad \bs q=\bs q_{\rm BZ}+\bs G,
 \label{eq:projected-basis-form-factor}
\end{equation}
where $\bs q$ is the full momentum transfer, with reduced part $\bs q_{\rm BZ}$ and reciprocal vector $\bs G$. It vanishes unless $\bs k_1\equiv\bs k_3+\bs q$.

Electrons interact through the Coulomb interaction screened by two symmetric metallic gates,
\begin{equation}
 V_q=\frac{e^2}{2\epsilon_0\epsilon_r}\frac{\tanh(qd_s)}q,\qquad q=|\bs q|,
\end{equation}
with gate distance $d_s=25\,\mathrm{nm}$ and relative permittivity $\epsilon_r=15$. In the projected basis, the direct and antisymmetrized matrix elements are
\begin{equation}
 V_{\mathbf{1}\mathbf{2};\mathbf{3}\mathbf{4}}
 =\frac1{A_{\rm uc}}\sum_{\bs q}
      V_q\,\mathcal F_{\mathbf{1}\mathbf{3}}(\bs q)
         \mathcal F_{\mathbf{2}\mathbf{4}}(-\bs q),\qquad
 \overline V_{\mathbf{1}\mathbf{2};\mathbf{3}\mathbf{4}}
 =V_{\mathbf{1}\mathbf{2};\mathbf{3}\mathbf{4}}
  -V_{\mathbf{1}\mathbf{2};\mathbf{4}\mathbf{3}}.
 \label{eq:projected-coulomb}
\end{equation}
In the direct term, electron $\mathbf3$ scatters into $\mathbf1$ and electron $\mathbf4$ into $\mathbf2$. $\overline V$ subtracts the exchange process. The projected Hamiltonian is then
\begin{equation}
 \hat{H}=\sum_{\mathbf{1}\mathbf{2}}h^0_{\mathbf{1}\mathbf{2}}
               c_{\mathbf{1}}^\dagger c_{\mathbf{2}}
       +\frac1{4N}\sum_{\mathbf{1}\mathbf{2}\mathbf{3}\mathbf{4}}
               \overline V_{\mathbf{1}\mathbf{2};\mathbf{3}\mathbf{4}}\,
               c_{\mathbf{1}}^\dagger c_{\mathbf{2}}^\dagger c_{\mathbf{4}}c_{\mathbf{3}},
 \label{eq:projected-hamiltonian}
\end{equation}
which is Eq.~(1) of the main text restricted to the retained states. The retained conduction states are empty at charge neutrality.

%==============================================================================
\section{Hartree--Fock calculation and self-doping}
%==============================================================================

\subsection{Self-consistent HF state}

We solve the HF equations in the projected basis. Crystal translations make the one-body density matrix block diagonal in reduced momentum, with arbitrary mixing among the seven states at each $\bs k$,
\begin{equation}
 \mathcal{P}_{ab}(\bs k)=\langle c^\dagger_{\bs k a}c_{\bs k b}\rangle,\qquad
 \mathcal{P}_{\mathbf{1}\mathbf{2}}=\delta_{\bs k_1\bs k_2}\mathcal{P}_{a_1a_2}(\bs k_1).
\end{equation}
With this convention, the HF Hamiltonian, containing both Hartree and Fock terms, is
\begin{equation}
 h^{\HF}_{\mathbf{1}\mathbf{2}}[\mathcal{P}]
   =h^0_{\mathbf{1}\mathbf{2}}+\frac1N\sum_{\mathbf{3}\mathbf{4}}
      \overline V_{\mathbf{1}\mathbf{3};\mathbf{2}\mathbf{4}}
      \mathcal{P}_{\mathbf{3}\mathbf{4}}.
 \label{eq:hf-self-consistency}
\end{equation}
At each iteration, we diagonalize $h^\HF(\bs k)$, with eigenstates $|\bs k,\alpha\rangle_{\HF}$ and energies $\varepsilon^{\HF}_\alpha(\bs k)$ ordered by the HF band index $\alpha=0,\ldots,6$. We occupy the $N_e$ lowest states across all momenta and bands and update the density matrix using the optimal damping algorithm~\cite{Cances2000} until it is self-consistent to within $10^{-12}$. Initial states are random or are transferred from converged solutions at other particle numbers, and no rotational symmetry is imposed. The converged state has occupations $n_{\bs k\alpha}\in\{0,1\}$, may contain partially filled bands, and has the density matrix
\begin{equation}
 \mathcal{P}_{ab}(\bs k)=\sum_\alpha n_{\bs k\alpha}\,
 {}_{\HF}\langle\bs k,\alpha|\bs k,a\rangle
 \langle\bs k,b|\bs k,\alpha\rangle_{\HF}.
 \label{eq:hf-eigenstates}
\end{equation}
The operator $\psi^\dagger_{\bs k\alpha}=\sum_a\langle\bs k,a|\bs k,\alpha\rangle_{\HF}c^\dagger_{\bs k a}$ creates an electron at $\bs{k}$ in HF band $\alpha$. At $n_e=0.4375\times10^{12}\,\mathrm{cm}^{-2}$, where we analyze pairing in most detail, the converged charge density is threefold symmetric to a relative accuracy of $3\times10^{-5}$, and the two lowest bands to $2\times10^{-4}\,\mathrm{meV}$.

The fluctuation and pairing calculations use the Coulomb interaction in the HF eigenbasis. Moving forward, we will then use bold labels to denote HF states, $\mathbf{1}=(\bs k_1,\alpha_1)$. The form factors in the HF basis $\mathcal F^{\HF}_{\mathbf{1}\mathbf{3}}(\bs q)={}_{\HF}\langle\mathbf{1}|e^{i\bs q\cdot\hat{\bs r}}|\mathbf{3}\rangle_{\HF}$ define $V^\HF$ and $\overline V^\HF$ as in Eq.~\eqref{eq:projected-coulomb}. Hermiticity and Fermi statistics imply
\begin{equation}
 \big(\overline V^\HF_{\mathbf1\mathbf2;\mathbf3\mathbf4}\big)^*=\overline V^\HF_{\mathbf3\mathbf4;\mathbf1\mathbf2},\qquad
 \overline V^\HF_{\mathbf1\mathbf2;\mathbf3\mathbf4}=-\overline V^\HF_{\mathbf2\mathbf1;\mathbf3\mathbf4}=-\overline V^\HF_{\mathbf1\mathbf2;\mathbf4\mathbf3}.
 \label{eq:hf-interaction}
\end{equation}

\subsection{Energy and self-doping}

The HF energy is
\begin{equation}
 E_{\HF}[\mathcal{P}]=\frac12\operatorname{Tr}
   \left[(h^0+h^{\HF}[\mathcal{P}])\mathcal{P}^T\right],
\end{equation}
where the factor $1/2$ corrects for double counting the interaction. A crystal with $N$ cells and $N_e$ electrons has density $n_e=N_e/(NA_{\rm uc})$, and we define the self-doping as the departure from one electron per cell,
\begin{equation}
 \delta n=n_e-\frac1{A_{\rm uc}}=n_e\left(1-\frac{N}{N_e}\right).
 \label{eq:hf-energy-self-doping}
\end{equation}
To vary $\delta n$ at fixed density, we keep the mesh, and hence $N$, fixed and change $N_e$, which sets the crystal period $L$. For each of the thirteen densities studied, we scan particle numbers, including the commensurate one, $N_e=N$, within $|\delta n|\le0.25\times10^{12}\,\mathrm{cm}^{-2}$ and report the optimal doping
\begin{equation}
 \delta n^\star=\underset{\delta n}{\arg\min}
     \left.\frac{E_{\HF}}{N_e}\right|_{n_e}.
\end{equation}

%==============================================================================
\section{Time-dependent Hartree--Fock calculation}
\label{sec:tdhf}
%==============================================================================

\begin{figure}[t]
 \centering
 \includegraphics[width=\linewidth]{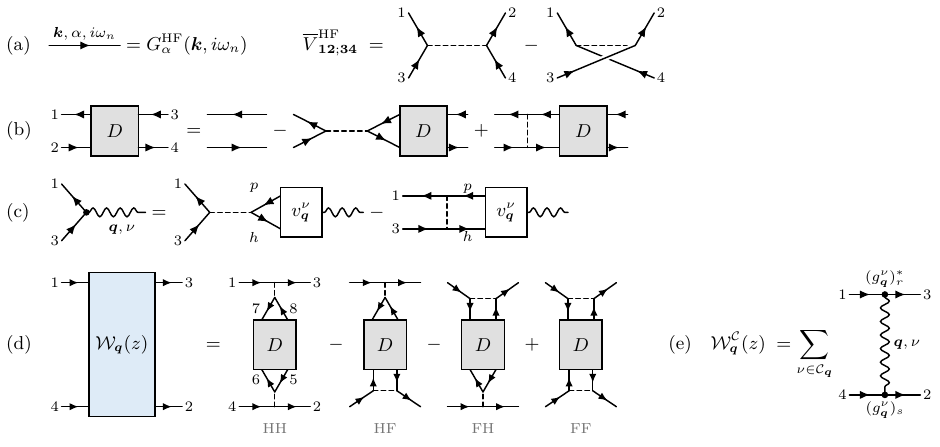}
 \caption{
 \textbf{Diagrams for the TDHF response and the mediated interaction.}
 (a) Solid lines are HF propagators and dashed lines the Coulomb interaction. The antisymmetrized vertex is the direct term minus the exchange term. (b)~Bethe--Salpeter equation~\eqref{eq:bse}, the direct part of the kernel inserts Hartree bubbles and the exchange part Fock ladders. (c)~Electron--mode coupling~\eqref{eq:electron-mode-coupling} given by the Hartree minus the Fock attachment of the TDHF eigenvector $v^\nu_{\bs q}$ (white box). (d) Mediated interaction $\mathcal{W}_{\bs{q}}(z)$~\eqref{eq:mediated} (light-blue block). The four Hartree and Fock attachments of $D$, with ports labeled as in Eq.~\eqref{eq:mediated}. (e)~Canonical part~\eqref{eq:mode-exchange}, where electrons exchange TDHF eingenmodes (wavy line, $\eta^\nu_{\bs q}/(\Omega^\nu_{\bs q}-z)$). %External legs carry HF bands 0 and 1, and internal lines all seven bands. 
 }
 \label{fig:diagrams}
\end{figure}

\subsection{HF reference action and Bethe--Salpeter equation}

We describe neutral fluctuations about the HF state using the TDHF approximation. We describe the approximation following the presentation of Ref.~\cite{vasiliou2024electrons} for later convenience. We start by writing the partition function as a path integral over the HF Slater determinant, so that the interaction is normal-ordered relative to the HF state. Up to a constant, the HF action is
\begin{equation}
 \mathcal S=\int_0^\beta d\tau\,
 \Big[\sum_{\mathbf{1}}\bar\psi_{\mathbf{1}}(\partial_\tau+\xi_{\mathbf{1}})\psi_{\mathbf{1}}
 +\frac{1}{4N}\sum_{\mathbf{1}\mathbf{2}\mathbf{3}\mathbf{4}}\overline V^{\HF}_{\mathbf{1}\mathbf{2};\mathbf{3}\mathbf{4}}
 :\bar\psi_{\mathbf{1}}\bar\psi_{\mathbf{2}}\psi_{\mathbf{4}}\psi_{\mathbf{3}}:_{\HF}\Big],
 \label{eq:hf-action}
\end{equation}
where $\xi_\alpha(\bs k)=\varepsilon^{\HF}_\alpha(\bs k)-\mu$ such that the bare propagator is the HF propagator $G^{\HF}_\alpha(\bs k,i\omega_n)=[i\omega_n-\xi_\alpha(\bs k)]^{-1}$.

Time-dependent Hartree--Fock (TDHF) describes small oscillations of the Slater determinant about the HF state. Diagrammatically, it is the conserving approximation~\cite{baym1961conservation} whose particle--hole irreducible vertex is the antisymmetrized bare interaction, i.e., the functional derivative of the HF self-energy. The particle--hole propagator $D$ therefore sums all chains of Hartree (bubble) and Fock (ladder) insertions, including mixed sequences. It satisfies the Bethe--Salpeter equation (BSE) showin in Fig.~\ref{fig:diagrams}(b)~\cite{vasiliou2024electrons},
\begin{equation}
 D_{\bs q}(z)=D^0_{\bs q}(z)-D^0_{\bs q}(z)\mathcal K_{\bs q}D_{\bs q}(z),\qquad
 D^0_{\mathbf{1}\mathbf{2};\mathbf{3}\mathbf{4}}(z)=\delta_{\mathbf{1}\mathbf{3}}\delta_{\mathbf{2}\mathbf{4}}
 \frac{n_{\mathbf{2}}-n_{\mathbf{1}}}{\varepsilon^{\HF}_{\mathbf{1}}-\varepsilon^{\HF}_{\mathbf{2}}-z},\qquad
 \mathcal K_{\mathbf{1}\mathbf{2};\mathbf{3}\mathbf{4}}=\frac{\overline V^{\HF}_{\mathbf{1}\mathbf{4};\mathbf{2}\mathbf{3}}}{N}.
 \label{eq:bse}
\end{equation}
A pair label $(\mathbf1,\mathbf2)$ refers to the bilinear $\psi^\dagger_{\mathbf1}\psi_{\mathbf2}$ with reduced momentum $\bs q=\bs k_1-\bs k_2$, $z$ is a complex bosonic energy, and the occupations are those of the HF state. In the kernel, the direct part of $\overline V^\HF$ connects the two pair densities (bubble), whereas the exchange part connects the particle lines with each other and the hole lines with each other (ladder rung). The same equations follow from linearizing the TDHF equations of motion~\cite{thouless1961vibrational,thouless1962time,rowe1968equations}.

\subsection{Collective modes}
\label{sec:collective}

Because $D^0$ vanishes when $n_{\mathbf1}=n_{\mathbf2}$, the response lives in the space of pairs connecting an occupied state $h$ to an empty state $p$. At momentum $\bs q$, the forward sector contains the pairs $(p,h)$ with $(h,p)\in\mathcal X_{\bs q}$, and the backward sector the reversed pairs $(h',p')\in\mathcal X_{-\bs q}$, where
\begin{equation}
 \mathcal X_{\bs q}=\{(h,p):n_h=1,\ n_p=0,\ \bs k_p-\bs k_h=\bs q\}.
\end{equation}
On these sectors $D^0_{\bs q}(z)$ equals $(\varepsilon^\HF_p-\varepsilon^\HF_h-z)^{-1}$ and $(\varepsilon^\HF_{p'}-\varepsilon^\HF_{h'}+z)^{-1}$, respectively. Hence $(D^0_{\bs q})^{-1}=E_{\bs q}-zS_{\bs q}$, where $E_{\bs q}$ contains the positive pair energies and $S_{\bs q}=\operatorname{diag}(\openone,-\openone)$ is the particle--hole metric. Multiplying the BSE by $(D^0_{\bs q})^{-1}$ gives the TDHF response as a resolvent,
\begin{equation}
 D_{\bs q}(z)=\big(\mathcal H_{\bs q}-zS_{\bs q}\big)^{-1},\qquad
 \mathcal H_{\bs q}=E_{\bs q}+\mathcal K_{\bs q}=\begin{pmatrix}A_{\bs q}&B_{\bs q}\\
                              B_{\bs q}^{\dagger}&A_{-\bs q}^{*}\end{pmatrix},
 \label{eq:resolvent}
\end{equation}
with the familiar TDHF blocks
\begin{equation}
 (A_{\bs q})_{hp,h'p'}=
  (\varepsilon_p^{\HF}-\varepsilon_h^{\HF})\delta_{hh'}\delta_{pp'}
       +\frac{\overline V^{\HF}_{p h';h p'}}{N},\qquad
 (B_{\bs q})_{hp,h'p'}=\frac{\overline V^{\HF}_{p p';h h'}}{N}.
 \label{eq:tdhf-blocks}
\end{equation}
Both pairs belong to $\mathcal X_{\bs q}$ in $A_{\bs q}$, whereas $(h',p')\in\mathcal X_{-\bs q}$ in $B_{\bs q}$; the lower blocks follow from Eq.~\eqref{eq:hf-interaction}. The Hermitian matrix $\mathcal H_{\bs q}$ is the stability matrix of the HF state~\cite{Thouless1960}.

The poles of $D_{\bs q}(z)$ are the TDHF collective frequencies, which, for electron crystals, include phonons~\cite{cote1991collective,dong2025phonons}. They are given by the generalized eigenvalue problem of Eq.~(2) of the main text,
\begin{equation}
 \mathcal H_{\bs q}v^\nu_{\bs q}=\Omega^\nu_{\bs q}S_{\bs q}v^\nu_{\bs q},
 \qquad v^\nu_{\bs q}=\binom{X_{\bs q\nu}}{Y_{\bs q\nu}}.
 \label{eq:tdhf-modes}
\end{equation}
A complex frequency signals a dynamical instability of the HF state. We call canonical the real, nonzero-frequency modes with nonvanishing metric norm,
\begin{equation}
 \mathcal C_{\bs q}=\{\nu:\Omega^\nu_{\bs q}\in\mathbb R\setminus\{0\},\ (v^\nu_{\bs q})^\dagger S_{\bs q}v^\nu_{\bs q}\ne0\},
\end{equation}
and normalize them as $(v^\nu_{\bs q})^\dagger S_{\bs q}v^{\nu'}_{\bs q}=\eta^\nu_{\bs q}\delta_{\nu\nu'}$, with metric sign $\eta^\nu_{\bs q}=\pm1$. Particle--hole conjugation maps each canonical mode at $\bs q$ to a partner at $-\bs q$ with opposite frequency and metric sign. The positive-norm modes are created by the operators $Q^\dagger_{\nu\bs q}$ of the main text. When the canonical modes form a complete set, the normalization gives $V^{-1}=\eta V^\dagger S_{\bs q}$ for the matrix $V$ of eigenvectors, and writing $\mathcal H_{\bs q}-zS_{\bs q}=S_{\bs q}V(\Omega-z)V^{-1}$ yields
\begin{equation}
 D_{\bs q}(z)=\sum_{\nu\in\mathcal C_{\bs q}}
   \frac{\eta^\nu_{\bs q}\,v^\nu_{\bs q}(v^\nu_{\bs q})^\dagger}{\Omega^\nu_{\bs q}-z}.
 \label{eq:pole-representation}
\end{equation}
Each mode thus contributes a pole whose weight is set by its eigenvector and metric sign. Zero modes, defective eigenvalues, or complex frequencies make the canonical set incomplete, in which case the resolvent itself must be used.

\textit{Numerical evaluation.}---For the pairing calculations, we evaluate the resolvent in a reduced subspace of a few hundred dimensions, compared with about $1.4\times10^{4}$ particle--hole pairs. This subspace is invariant under $S_{\bs q}$ and contains the low-energy modes, along with response vectors generated by the electron attachments in Sec.~\ref{sec:interaction}. %The retained modes are preserved to numerical precision in the reduced problem. At four tested momenta at $n_e=0.4375\times10^{12}\,\mathrm{cm}^{-2}$, the tested response contractions agree with the full calculation within $0.05\%$ at zero frequency and $1\%$ at sampled imaginary frequencies through $10\,\mathrm{meV}$.

%\textit{Static response at $\Gamma$.}---In the translation-invariant limit, uniform translations of the crystal are zero modes at $\bs q=0$, and the static response $D_{\boldsymbol0}(0)$ is singular because a uniform force displaces the crystal freely. Numerical near-zero splittings are retained. For $D_{\boldsymbol0}(0)$ only, we fix the two uniform displacement coordinates, obtained from the commutators of the momentum operators with the HF density matrix, while keeping their conjugate momenta and all other coordinates. This fixes the crystal origin without removing the $\Gamma$ response. Nonzero frequencies use the unconstrained resolvent.

%==============================================================================
\section{Electron--boson coupling and effective retarded interaction}
\label{sec:interaction}
%==============================================================================

\subsection{Electron--mode coupling}

An external HF electron couples to a collective fluctuation through the Hartree and Fock fields that the fluctuation induces. Contracting the antisymmetrized interaction with the TDHF eigenvector gives the vertex of Fig.~\ref{fig:diagrams}(c)~\cite{vasiliou2024electrons},
\begin{equation}
 (g^\nu_{\bs q})_r
 =\frac1{\sqrt N}\Big[
 \sum_{(h,p)\in\mathcal X_{\bs q}}
       \overline V^{\HF}_{\mathbf{1}h;\mathbf{3}p}X^\nu_{\bs q,hp}
 +\sum_{(h',p')\in\mathcal X_{-\bs q}}
       \overline V^{\HF}_{\mathbf{1}p';\mathbf{3}h'}Y^\nu_{\bs q,h'p'}
 \Big],\qquad
 r=(\mathbf{3},\mathbf{1}),\quad \bs k_1-\bs k_3=\bs q,
 \label{eq:electron-mode-coupling}
\end{equation}
for an electron scattering from $\mathbf3$ into $\mathbf1$ while absorbing the mode. The direct part of $\overline V^\HF$ couples the electron density to the mode's pair density (Hartree), and the exchange part connects the external line to the pair's particle and hole lines (Fock). Both forward and backward amplitudes contribute. For our pairing calculation, external electrons are restricted to HF bands 0 and 1, while the mode retains its full seven-band structure.

\subsection{Effective retarded interaction}

Attaching the antisymmetrized interaction to both ends of the TDHF response gives the mediated interaction of Fig.~\ref{fig:diagrams}(d),
\begin{equation}
 [\mathcal{W}_{\bs q}(z)]_{rs}
 =\frac1N\sum_{\mathbf{5}\mathbf{6}\mathbf{7}\mathbf{8}}
       \overline V^{\HF}_{\mathbf{2}\mathbf{6};\mathbf{4}\mathbf{5}}
       [D_{\bs q}(z)]_{\mathbf{5}\mathbf{6};\mathbf{7}\mathbf{8}}
       \overline V^{\HF}_{\mathbf{3}\mathbf{7};\mathbf{1}\mathbf{8}},\qquad
 r=(\mathbf3,\mathbf1),\quad s=(\mathbf4,\mathbf2),
 \label{eq:mediated}
\end{equation}
with $\bs k_1-\bs k_3=\bs k_2-\bs k_4=\bs q$ and external bands in $\{0,1\}$. The internal pairs run over both the forward and backward sectors. Electron $\mathbf1$ scatters into $\mathbf3$ at one end, and electron $\mathbf4$ into $\mathbf2$ at the other, so the momentum $\bs q$ and energy $z$ are transferred from the first electron to the second. Expanding each $\overline V^\HF$ into its direct and exchange parts yields the four attachments shown in Fig.~\ref{fig:diagrams}(d).

Inserting the pole representation~\eqref{eq:pole-representation} turns each end into an electron--mode coupling. The end attached to $\mathbf4\to\mathbf2$ is $(g^\nu_{\bs q})_s$, and Hermiticity of $\overline V^\HF$ shows that the end attached to $\mathbf1\to\mathbf3$ is $(g^\nu_{\bs q})^*_r$. The canonical part of $\mathcal W$ is therefore
\begin{equation}
 [\mathcal{W}^{\mathcal C}_{\bs q}(z)]_{rs}
 =\sum_{\nu\in\mathcal C_{\bs q}}
       \frac{\eta^\nu_{\bs q}\,(g^\nu_{\bs q})_s
                      (g^\nu_{\bs q})_r^*}
            {\Omega^\nu_{\bs q}-z},
 \label{eq:mode-exchange}
\end{equation}
the mode-exchange form of Ref.~\cite{vasiliou2024electrons}, shown in Fig.~\ref{fig:diagrams}(e). The pairing calculation uses the full $\mathcal W$ of Eq.~\eqref{eq:mediated}. $\mathcal W^{\mathcal C}$ equals it only when the canonical set is complete, and serves to interpret it and to isolate selected modes. Its residues carry band and momentum indices, and the spectra at $\bs q$ and $-\bs q$ need not coincide, so $\mathcal W$ is not in general a scalar attractive interaction.

Overall, $\mathcal W$ contains all TDHF bubble and ladder chains for the seven-band model, treating collective modes and the particle--hole continuum on equal footing. It is the electron-crystal analog of the fluctuation-exchange interactions used to describe pairing mediated by spin and charge fluctuations~\cite{berk1966effect,scalapino1986d,bickers1989conserving,graser2009near,scalapino2012common}. It omits self-energy corrections beyond HF, vertex corrections beyond TDHF, and particle--particle fluctuations. %We also work with HF electrons in the laboratory frame. Long-wavelength Goldstone modes need not decouple from the electrons~\cite{watanabe2014criterion}. In particular, Vasiliou \textit{et al.}~\cite{vasiliou2024electrons} showed that their interband couplings need not vanish in this frame and are modified in a frame that follows the local order parameter; we do not performed this transformation for crystal translations.

%==============================================================================
\section{Finite-momentum pairing}
\label{sec:pairing}
%==============================================================================

\subsection{Cooper vertex and Pauli exchange}

\begin{figure}[t]
 \centering
 \includegraphics[width=0.92\linewidth]{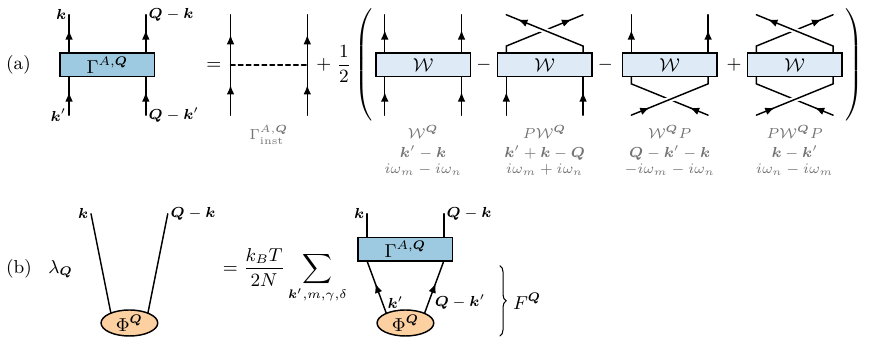}
 \caption{\textbf{Pairing calculation.} (a)~Pairing vertex~\eqref{eq:cooper-vertex}. The incoming pair $(\bs k',\gamma,i\omega_m;\,\bs Q-\bs k',\delta,-i\omega_m)$ at the bottom scatters into the outgoing pair $(\bs k,\alpha,i\omega_n;\,\bs Q-\bs k,\beta,-i\omega_n)$ at the top. The instantaneous part is the bare antisymmetrized interaction (dashed line). The mediated part contains the full interaction $\mathcal W$ of Fig.~\ref{fig:diagrams}(d) (light-blue block) in the four routings generated by the Pauli projectors, where $P$ exchanges the outgoing or incoming legs. Below each routing are the momentum and energy transferred from the left to the right electron. (b)~Linearized Eliashberg equation~\eqref{eq:gap-equation}. The anomalous self-energy $\Phi^{\bs Q}$ and the two HF propagators form the pair response $F^{\bs Q}$, which $\Gamma^{A,\bs Q}$ scatters back into $\Phi^{\bs Q}$. Legs without arrows are amputated.}
 \label{fig:pairing}
\end{figure}

We consider pairing within the occupied spin and valley sector, with total crystal momentum $\bs Q$. 
%Because the displacement field and valley polarization break inversion and time-reversal symmetry, the leading instability need not occur at $\bs Q=0$~\cite{fischer2018superconductivity}. 
The pairing vertex scatters an incoming pair $(\bs k',\gamma,i\omega_m;\,\bs Q-\bs k',\delta,-i\omega_m)$ into an outgoing pair $(\bs k,\alpha,i\omega_n;\,\bs Q-\bs k,\beta,-i\omega_n)$, where $\alpha,\beta,\gamma,\delta\in\{0,1\}$ are HF bands. Its instantaneous part is the bare antisymmetrized interaction, with the sign convention $\Gamma=-V_{\rm eff}$,
\begin{equation}
 [\Gamma^{A,\bs Q}_{\mathrm{inst}}]_{\alpha\beta;\gamma\delta}(\bs k,\bs k')
   =-\overline V^{\HF}_{(\bs k,\alpha)(\bs Q-\bs k,\beta);
                         (\bs k',\gamma)(\bs Q-\bs k',\delta)}.
\end{equation}
The retarded part follows from Eq.~\eqref{eq:mediated} with
\begin{equation}
 \begin{gathered}
 \bs q=\bs k'-\bs k,\qquad z=i\omega_m-i\omega_n,\qquad
 r=(\bs k,\alpha;\bs k',\gamma),\qquad
 s=(\bs Q-\bs k',\delta;\bs Q-\bs k,\beta),\\
 [\mathcal{W}^{\bs Q}]_{\alpha\beta;\gamma\delta}(\bs k,i\omega_n;\bs k',i\omega_m)=[\mathcal{W}_{\bs q}(z)]_{rs},
 \end{gathered}
 \label{eq:cooper-routing}
\end{equation}
which corresponds to the first mediated diagram in Fig.~\ref{fig:pairing}(a).

Fermi statistics require antisymmetry under the simultaneous exchange of the two electrons' bands, momenta, and frequencies. We denote this exchange by $P$,
\begin{equation}
 (P\Phi^{\bs Q})_{\alpha\beta}(\bs k,i\omega_n)=\Phi^{\bs Q}_{\beta\alpha}(\bs Q-\bs k,-i\omega_n),\qquad
 \Pi^{\bs Q}_-=\tfrac12\big(1-P\big),
\end{equation}
where $\omega_{-n-1}=-\omega_n$. Since $P^2=1$, $\Pi^{\bs Q}_-=\tfrac12(1-P)$ is the projector onto pair functions that are antisymmetric under the exchange of the two electrons, $P\Phi^{\bs Q}=-\Phi^{\bs Q}$, i.e., onto the Pauli-allowed pair states. Acting on a kernel, $P$ exchanges either its outgoing or its incoming labels. The bare tensor $\Gamma^{A,\bs Q}_{\rm inst}$ is already antisymmetric by Eq.~\eqref{eq:hf-interaction}, whereas $\mathcal{W}^{\bs Q}$ is not. The full pairing vertex is
\begin{equation}
 \Gamma^{A,\bs Q}
 =\Gamma^{A,\bs Q}_{\mathrm{inst}}
   +2\Pi_-^{\bs Q}\mathcal{W}^{\bs Q}\Pi_-^{\bs Q},\qquad
 2\Pi_-^{\bs Q}\mathcal{W}^{\bs Q}\Pi_-^{\bs Q}=\tfrac12\big(\mathcal{W}^{\bs Q}-P\mathcal{W}^{\bs Q}-\mathcal{W}^{\bs Q}P+P\mathcal{W}^{\bs Q}P\big).
 \label{eq:cooper-vertex}
\end{equation}
The four terms are the routings of Fig.~\ref{fig:pairing}(a): $\mathcal W^{\bs Q}$ and $P\mathcal W^{\bs Q}P$ carry the direct momentum transfer $\pm(\bs k'-\bs k)$, whereas $P\mathcal W^{\bs Q}$ and $\mathcal W^{\bs Q}P$ carry the crossed transfer $\pm(\bs k'+\bs k-\bs Q)$, as for the two particle--hole channels of an antisymmetrized vertex~\cite{rohringer2018diagrammatic}. We retain all four.

We make a few clarifying comments about the normalization factor appearing in~\eqref{eq:cooper-vertex}. To combine the bare and mediated interactions, we write both in the same form. In the action, a two-body interaction enters as $(1/4N)\sum V^{\rm eff}_{\mathbf1\mathbf2;\mathbf3\mathbf4} \bar\psi_{\mathbf1}\bar\psi_{\mathbf2}\psi_{\mathbf4}\psi_{\mathbf3}$, with $V^{\rm eff}$ antisymmetric in both pairs, and the pairing vertex is $\Gamma=-V_{\rm eff}$. The bare interaction in Eq.~\eqref{eq:hf-action} already has this form, with $V_{\rm eff}=\overline V^\HF$. Integrating out the Gaussian TDHF fluctuations instead generates $-(1/2N)\sum\mathcal W\,\bar\psi\bar\psi\psi\psi$, written in the Cooper labels of Eq.~\eqref{eq:cooper-routing}, where $1/2$ is the usual Gaussian factor and $\mathcal W$ is not antisymmetric. Because the Grassmann product changes sign under exchange within either pair, only the antisymmetric part $\Pi_-\mathcal W\Pi_-$ contributes, so this term equals $(1/4N)\sum(-2\Pi_-\mathcal W\Pi_-)\,\bar\psi\bar\psi\psi\psi$. Hence $V_{\rm eff}=\overline V^\HF-2\Pi_-\mathcal W\Pi_-$, and $\Gamma=-V_{\rm eff}$ gives Eq.~\eqref{eq:cooper-vertex}. The factor of two is thus the ratio of the Gaussian factor $1/2$ to the antisymmetrized normalization $1/4$.

% \textit{Normalization.}---The relative factor of two follows from the effective action. The bare interaction enters Eq.~\eqref{eq:hf-action} as $(1/4N)\,\overline V^\HF\bar\psi\bar\psi\psi\psi$, already antisymmetric in both pairs. Integrating out the Gaussian TDHF fluctuations produces instead $-(1/2N)\,\mathcal{W}\bar\psi\bar\psi\psi\psi$, where $1/2$ is the Gaussian factor and $\mathcal{W}$ is not antisymmetrized. Since the Grassmann product is antisymmetric within each pair, only $2\Pi_-\mathcal W\Pi_-$ survives when this term is written in the normalization of the bare one. The effective interaction is therefore $V_{\rm eff}=\overline V^\HF-2\Pi_-\mathcal{W}\Pi_-$, which gives Eq.~\eqref{eq:cooper-vertex} \textcolor{red}{see how best to frame this}.

\subsection{Linearized pairing}

We perform a one-shot pairing calculation with fixed HF propagators and a fixed TDHF interaction, as in RPA-based studies of fluctuation-mediated pairing~\cite{scalapino1986d,graser2009near}. At each temperature, the chemical potential is set by the density using all seven HF bands,
\begin{equation}
 n_e A_{\rm uc}=\frac1{N}\sum_{\bs k}\sum_{\alpha=0}^{6}
           f_T[\varepsilon^{\HF}_\alpha(\bs k)-\mu],\qquad
 f_T(x)=\frac{1}{e^{x/(k_BT)}+1},
\end{equation}
while pairing is restricted to the two lowest bands. The anomalous self-energy $\Phi^{\bs Q}$, with $\Pi_-^{\bs Q}\Phi^{\bs Q}=\Phi^{\bs Q}$, generates the linear pair response
\begin{equation}
 F^{\bs Q}_{\alpha\beta}(\bs k,i\omega_n)
 =G^{\HF}_\alpha(\bs k,i\omega_n)
       \Phi^{\bs Q}_{\alpha\beta}(\bs k,i\omega_n)
       G^{\HF}_\beta(\bs Q-\bs k,-i\omega_n),
 \label{eq:pair-fields}
\end{equation}
and the linearized Eliashberg equation~\cite{eliashberg1960interactions,allen1983theory,carbotte1990properties,takimoto2004strong}, shown in Fig.~\ref{fig:pairing}(b), reads
\begin{equation}
 \lambda_{\bs Q}\Phi^{\bs Q}_{\alpha\beta}(\bs k,i\omega_n)
 =\frac{k_BT}{2N}
  \sum_{\bs k',m,\gamma,\delta}
    \Gamma^{A,\bs Q}_{\alpha\beta;\gamma\delta}
          (\bs k,i\omega_n;\bs k',i\omega_m)
    F^{\bs Q}_{\gamma\delta}(\bs k',i\omega_m),
 \label{eq:gap-equation}
\end{equation}
where the factor $1/2$ compensates for the two orderings of the Cooper labels. The kernel is generally non-Hermitian. A real eigenvalue that reaches unity marks the pairing instability in this approximation, and the corresponding eigenvector gives the structure of the incipient pair field. We compute the leading eigenvalues with ARPACK and report the largest one that is real to numerical precision, $|\operatorname{Im}\lambda|\le10^{-8}\max(1,|\lambda|)$.

The instantaneous interaction determines the constant high-frequency asymptote
$\Phi^{\boldsymbol{Q}}_{\infty}(\boldsymbol{k})
=\lim_{|\omega_n|\to\infty}
\Phi^{\boldsymbol{Q}}(\boldsymbol{k},i\omega_n)$,
since the retarded contribution vanishes in this limit. This asymptote satisfies
\begin{align}
\lambda_{\boldsymbol{Q}}
\Phi^{\boldsymbol{Q}}_{\infty,\alpha\beta}(\boldsymbol{k})
=
\frac{1}{2N}
\sum_{\boldsymbol{k}',\gamma,\delta}
[\Gamma^{A,\boldsymbol{Q}}_{\mathrm{inst}}]_{\alpha\beta;\gamma\delta}
(\boldsymbol{k},\boldsymbol{k}')
F^{\boldsymbol{Q},\mathrm{ET}}_{\gamma\delta}(\boldsymbol{k}'),
\end{align}
where
\begin{align}
F^{\boldsymbol{Q},\mathrm{ET}}_{\alpha\beta}(\boldsymbol{k})
=
k_BT\sum_{n\in\mathbb{Z}}
F^{\boldsymbol{Q}}_{\alpha\beta}(\boldsymbol{k},i\omega_n)
\end{align}
is the equal-time pair response. We solve the pairing equation within the frequency window \(-N_\omega\le n\le N_\omega-1\), together with the equation for \(\Phi^{\boldsymbol{Q}}_\infty\). Outside this window, we approximate the anomalous self-energy by its constant asymptote. The corresponding contribution to the equal-time response is then
\begin{align}
&k_BT\sum_{n\notin\mathrm{window}}
F^{\boldsymbol{Q}}_{\alpha\beta}(\boldsymbol{k},i\omega_n)
\simeq
\Phi^{\boldsymbol{Q}}_{\infty,\alpha\beta}(\boldsymbol{k})
\left[
\Lambda^{\boldsymbol{Q}}_{\alpha\beta}(\boldsymbol{k})
-k_BT\sum_{n\in\mathrm{window}}
G^{\mathrm{HF}}_\alpha(\boldsymbol{k},i\omega_n)
G^{\mathrm{HF}}_\beta(\boldsymbol{Q}-\boldsymbol{k},-i\omega_n)
\right],
\end{align}
where the full Matsubara sum has the closed form
\begin{align}
\Lambda^{\boldsymbol{Q}}_{\alpha\beta}(\boldsymbol{k})
&=
k_BT\sum_{n\in\mathbb{Z}}
G^{\mathrm{HF}}_\alpha(\boldsymbol{k},i\omega_n)
G^{\mathrm{HF}}_\beta(\boldsymbol{Q}-\boldsymbol{k},-i\omega_n)
=
\frac{
1-f_T[\xi_\alpha(\boldsymbol{k})]
-f_T[\xi_\beta(\boldsymbol{Q}-\boldsymbol{k})]
}{
\xi_\alpha(\boldsymbol{k})
+\xi_\beta(\boldsymbol{Q}-\boldsymbol{k})
}.
\end{align}
The last expression approaches \(-f_T'[\xi_\alpha(\boldsymbol{k})]\) when its denominator tends to zero. We include this analytic tail in the equal-time response and the instantaneous contribution to the pairing equation. The retarded convolution is evaluated only within the retained frequency window. Its out-of-window contribution is neglected as a separate numerical approximation and is not restored by the analytic tail above. At \(T=1\,\mathrm{K}\), the choice \(N_\omega=18\) corresponds to a maximum retained Matsubara energy of approximately \(9.48\,\mathrm{meV}\). 
%, except the $\bs Q=0$ points at $n_e=0.3750$, $0.4000$, $0.4125$, and $0.4250\times10^{12}\,\mathrm{cm}^{-2}$ in Fig.~3(a), which use $N_\omega=9$. Doubling $N_\omega$ from 9 to 18 lowers $\lambda_{\bs Q=0}$ by about $6\%$ at $n_e=0.3875$ and $1\%$ at $0.4375\times10^{12}\,\mathrm{cm}^{-2}$. The upturn at the largest density persists when the densities are compared at a common frequency window.

Pairing uses the HF mesh, so $\bs Q$ is a mesh momentum. We scan $\bs Q$ at $n_e=0.4375\times10^{12}\,\mathrm{cm}^{-2}$ on every third mesh point, reduced by threefold rotation, and refine the scan where $\lambda_{\bs Q}$ is largest. This produced the 91-momentum scan in the inset of Fig.~3(a). %The eight orbits checked at multiple momenta agree to about $0.04\%$. 
The largest eigenvalue among the sampled momenta occurs at $\bs Q_*=(-15\bs b_1+10\bs b_2)/36$, and we use these reduced coordinates at all densities. 
%Only one member of the maximizing orbit was evaluated directly.

\textit{Soft-mode test.}---For Fig.~4(d), the soft sector consists of the lowest positive-frequency, positive-norm TDHF mode below $0.1\,\mathrm{meV}$ at each momentum in the three finite-momentum pockets, excluding the low-energy region connected to $\Gamma$, together with the negative-frequency partners of these 60 modes at $-\bs q$. The selected poles and residues are evaluated from the same reduced response used to construct $\mathcal W$. Their pole terms in Eq.~\eqref{eq:mode-exchange} define $\mathcal W_{\rm soft}$. The ``full\,$-$\,soft'' kernel replaces $\mathcal W$ by $\mathcal W-\mathcal W_{\rm soft}$ at all frequencies, including the static one, and the ``instantaneous\,$+$\,soft'' kernel keeps only $\Gamma^{A,\bs Q}_{\rm inst}$ and $2\Pi_-\mathcal W_{\rm soft}\Pi_-$. All three kernels are solved at the same $T$, $\bs Q_*$, and $N_\omega$. This test isolates the role of the finite-momentum soft modes at fixed $\bs Q_*$; it does not address the long-wavelength frame question of Sec.~\ref{sec:interaction}.

\bibliography{ref}